\documentstyle[preprint,aps,prd]{revtex}

\begin{document}
\draft
\tightenlines
\preprint{SNUTP-98-074}
\title{Perturbation Method beyond the Variational
Gaussian Approximation: The Liouville-Neumann Approach}

\author{Dongsu Bak, $^{1,}$\footnote{Electronic address:
dsbak@mach.uos.ac.kr}
Sang Pyo Kim,$^{2,}$\footnote{Electronic address: sangkim@knusun1.kunsan.ac.kr}
Sung Ku Kim, $^{3,}$\footnote{Electronic address: skkim@theory.ewha.ac.kr}
Kwang-Sup Soh, $^{4,}$\footnote{Electronic address: kssoh@phya.snu.ac.kr}
and Jae Hyung Yee $^{5,}$\footnote{Electronic address: jhyee@phya.yonsei.ac.kr}}

\address{$^1$ Department of Physics,
University of Seoul,
Seoul 130-743, Korea \\
$^2$ Department of Physics,
Kunsan National University,
Kunsan 573-701, Korea \\
$^3$ Department of Physics and Research Institute for Basic Sciences,
Ewha Womans University,
Seoul 120-750, Korea\\
$^4$ Department of Physics Education,
Seoul National University,
Seoul 151-742, Korea\\
$^5$ Department of Physics,
Yonsei University,
Seoul 120-749, Korea}

\date{\today}
\maketitle
\begin{abstract}
We have developed a variational perturbation theory based on the 
Liouville-Neumann equation, which enables one to systematically compute 
the perturbative correction terms to the variationally determined wave 
functions of the time-dependent systems. We then apply
the method to the time-independent anharmonic oscillator, and show that
the results agree with those of other variational perturbation theories.
We also show that the system has an interesting algebraic structure at 
the first order correction level.
\end{abstract}

\section{Introduction}

Recently considerable efforts have been devoted to construct
systematic procedures to improve the variational approximation
method of quantum mechanics and quantum field theory \cite{dineykhan,cea,lee}.
The so-called variational perturbation theory \cite{dineykhan,cea} is based
on the observation that the variationally determined approximate
ground state of an interacting system is the exact ground state of a new system
described by a quadratic (or, a soluble) Hamiltonian, and one uses the exact
eigenstates of this new Hamiltonian to formulate a perturbation
theory. Although this method is conceptually simple, the computation
of the perturbation series is quite complicated for complex
and especially time-dependent systems. The method of reference
\cite{lee}, on the other hand, is based on the background field
method, and provides a fairly simple procedure to compute the perturbative
correction terms to the variationally determined
Gaussian effective action. For time-independent systems, the latter
method gives the same effective potentials as those of
the variational perturbation theory of ref.\cite{dineykhan,cea}. If one is interested in the 
explicit time evolution of a system of interest, however, one must construct
explicit time-dependent wave function(al)s of the system,
and the methods of ref.\cite{dineykhan,cea,lee} are not of much help for this purpose.

It is the purpose of this paper to develop another method
of perturbative expansion around the variational approximation,
which provides a systematic procedure to evaluate the correction
terms to the variationally determined wave function(al)s of general 
time-dependent systems.

To do this we use the technique of Lewis and Riesenfeld \cite{lewis},
which is originally developed for solving the Schr\"{o}dinger equation for a
time-dependent harmonic oscillator. Let us consider the Schr\"{o}dinger equation
(unit $\hbar =1$),
\begin{equation}
i \frac{\partial}{\partial t} \vert \Psi \rangle
= \hat{H} (t) \vert \Psi \rangle,
\label{sch eq}
\end{equation}
where the Hamiltonian $\hat{H}$ may have explicit time-dependence. 
The method of ref.\cite{lewis} consists of introducing a non-degenerate Hermitian 
invariant operator, $\hat{I}$, which satisfies the Liouville-Neumann equation,
\begin{equation}
i \frac{\partial}{\partial t} \hat{I} (t) + [ \hat{I} (t), \hat{H} ] = 0.
\label{ln eq}
\end{equation}
It is the property of this invariant that the eigenvalues
are time-independent, $\frac{\partial e_n}{\partial t} = 0$ \cite{lewis}:
\begin{equation}
\hat{I} (t) \vert e_n, t \rangle = e_n \vert e_n, t \rangle.
\label{eigen}
\end{equation}
One can then show that the solution of the Schr\"{o}dinger equation (\ref{sch eq}) 
is given by 
\begin{equation}
\vert \Psi (t) \rangle = \sum_{n} C_n \vert e_n , t \rangle e^{i \int dt 
\langle e_n, t \vert i \frac{\partial}{\partial t} - \hat{H} \vert e_n, t \rangle }.
\label{sch eq2}
\end{equation}
Therefore the task of finding the solution of the Schr\"{o}dinger equation 
is equivalent to finding the invariant operators.

In oder to develop a new perturbative method based on the Lewis-Riesenfeld technique,
we will use a quantum mechanical system described by the Hamiltonian (\ref{anhar}) below as a 
simple example. For this purpose it is useful to express
the invariant operator, $\hat{I}$, in terms of lowering and
raising type operators $\hat{a}$ and $\hat{a}^\dagger$:
\begin{equation}
\hat{I} (t) = \sum_{n, r, s} \lambda^{n} b^{(r,s)}_{(n)} \hat{a}^{\dagger r} \hat{a}^{s} ,
\label{eq hati}
\end{equation}
where $\hat{a}$ and $\hat{a}^{\dagger}$ also satisfy
the Liouville-Neumann equation (\ref{ln eq}) for a new quadratic  
Hamiltonian $\hat{H}_G$ and the commutation relation,
\begin{equation}
[ \hat{a}, \hat{a}^{\dagger} ] = 1,
\label{com rel}
\end{equation}
and $\lambda$ is an appropriate expansion parameter.
The quadratic Hamiltonian $\hat{H}_G$ will be determined such that it gives
the variational approximation of the interacting system.
Now finding the invariant operator $\hat{I}$, which
enables one to obtain the solution to the Schr\"{o}dinger
equation (\ref{sch eq}), amounts to finding the annihilation and
creation operators which satisfy both the Liouville-Neumann
equation (\ref{ln eq}) for $\hat{H}_G$ and the normalization condition
(\ref{com rel}), and to finding the coefficients $b^{(r,s)}_{(n)}$ of Eq. (\ref{eq hati}).

To obtain the variational approximation method for an interacting
system we write $\hat{a}$ and $\hat{a}^{\dagger}$ in terms of
the dynamical variables of the system with an arbitrary complex
function introduced, and express the Hamiltonian in terms of these
$\hat{a}$ and $\hat{a}^{\dagger}$. We then truncate the
Hamiltonian up to quadratic terms in $\hat{a}$ and $\hat{a}^{\dagger}$
for an approximation. The requirement that $\hat{a}$ and $\hat{a}^{\dagger}$
satisfy the Liouville-Neumann equation with the truncated
quadratic Hamiltonian determines the equation for the
variational approximation of the original system.
We then treat the remaining part of the Hamiltonian
perturbatively to determine the annihilation and creation
operators of the interacting system as a perturbation
series, which leads to the Liouville-Neumann approach to the
variational perturbation theory.

In the next section we establish the variational approximation
method based on the Liouville-Neumann equation, by constructing
a quadratic Hamiltonian which determines the variational approximation of the 
interacting system. In section III we develop a systematic procedure to 
evaluate perturbative correction terms to the variationally determined wave 
function of the interacting system.
In section IV we apply the method to the case of the eigenvalue problem for 
the time-independent anharmonic 
oscillator, and show that the result is equivalent to that obtained by other 
variational perturbation theories. We conclude with some discussions on our 
method in the last section.

\section{Liouville-Neumann Approach to the Variational
Approximation}

To develop the variational perturbative method based on the
Liouville-Neumann equation, we consider the time-dependent anharmonic oscillator
described by the Hamiltonian,
\begin{equation}
\hat{H} = \frac{1}{2}\hat{p}^2 + \frac{{\omega}^2 (t)}{2} \hat{q}^2
+ \frac{\tilde{\lambda}(t)}{4} \hat{q}^4,
\label{anhar}
\end{equation}
where the mass is scaled to unity for simplicity, and $\omega$ and $\tilde{\lambda}$ are 
time dependent in general. 
It is convenient to express the time-dependent coupling constant $\tilde{\lambda}$ as 
\begin{equation}
\tilde{\lambda}(t) = \lambda \epsilon(t),
\label{timedependentlambdaepsilon}
\end{equation}
where $\lambda$ is the time-independent expansion parameter and $\epsilon(t)$ carries all 
the time-dependence of the coupling constant $\tilde{\lambda}(t)$. For the time-independent 
systems we will set $\epsilon(t) = 1$.
As explained in the previous section, 
the variationally approximated states of this system is the exact states
of a quadratic variational Hamiltonian.
To find such a quadratic Hamiltonian for the anharmonic oscillator
(\ref{anhar}), we introduce a set of annihilation and creation
operators \cite{kim},
\begin{eqnarray}
&{}\hat{a} = i \Bigl( u^* (t) \hat{p}
- \dot{u}^* (t) \hat{q} \Bigr),
\nonumber\\
&{}\hat{a}^{\dagger} = - i \Bigl(u (t) \hat{p}
- \dot{u} (t) \hat{q} \Bigr),
\label{cr-an}
\end{eqnarray}
where $u(t)$ is an arbitrary complex function satisfying the normalization
condition,
\begin{equation}
\bigl( \dot{u}^* u - \dot{u} u^* \bigr) = i,
\label{wronskian}
\end{equation}
due to the commutation relation (\ref{com rel}).

We now express the Hamiltonian (\ref{anhar}) in terms of $\hat{a}$
and $\hat{a}^{\dagger}$ of (\ref{cr-an}), which, after normal
ordering, becomes,
\begin{equation}
\hat{H} = \hat{H}_G + \lambda \hat{H}',
\end{equation}
where
\begin{eqnarray}
\hat{H}_G &=& \Bigl[ \dot{u}^* \dot{u} + \omega^2 u^* u + 3\lambda \epsilon 
(u^* u)^2 \Bigr] \Bigl( \hat{a}^{\dagger} \hat{a}
+ \frac{1}{2} \Bigr) - \frac{3}{4} \lambda \epsilon (u^* u)^2
\nonumber\\
 && + \frac{1}{2} \Bigl[\dot{u}^{*2} + \omega^2 u^{*2}
+ 3 \lambda \epsilon (u^* u) u^{*2} \Bigr] \hat{a}^{\dagger 2}
\nonumber\\
 && + \frac{1}{2} \Bigl[\dot{u}^{2} + \omega^2 u^{2}
+ 3\lambda \epsilon (u^* u) u^{2} \Bigr] \hat{a}^{ 2}
\label{op rep}
\end{eqnarray}
and
\begin{equation}
\hat{H}' = \frac{\epsilon}{4} \sum_{k = 0}^{4} {4 \choose k}
u^{* (4 - k)} (u)^k \hat{a}^{\dagger (4 - k)} \hat{a}^k.
\end{equation}
To find the quadratic Hamiltonian that determines the variational
approximation, we require the annihilation and creation operators
(\ref{cr-an}) to satisfy the Liouville-Neumann equation,
\begin{eqnarray}
i \frac{\partial}{\partial t} \hat{a}
+ [\hat{a}, \hat{H}_G ]
= 0,
\nonumber\\
i \frac{\partial}{\partial t} \hat{a}^{\dagger}
+ [\hat{a}^{\dagger}, \hat{H}_G]
= 0,
\label{lowest}
\end{eqnarray}
with the truncated Hamiltonian $\hat{H}_G$ of Eq.
(\ref{op rep}). Eq. (\ref{lowest}) requires the complex
function $u(t)$ of (\ref{cr-an}) and (\ref{wronskian})
to satisfy the equation of motion,
\begin{equation}
\ddot{u} + \omega^2 u + 3 \lambda \epsilon (u^* u) u  = 0.
\label{mean eq}
\end{equation}
Note that Eq. (\ref{mean eq}) is exactly the gap equation
appearing in the Gaussian approximation where the trial wave
functional is written as \cite{eboli},
\begin{equation}
\Psi(q) = N \exp \Bigl[i \frac{\dot{u}^*}{2 u^*} q^2 \Bigr].
\label{Psi eq10}
\end{equation}

To obtain solutions of the Schr\"{o}dinger equation with $\hat{H}_G$ 
one needs to find the eigenstates of the invariant operator 
$\hat{I} = \hat{a}^{\dagger} \hat{a}$, which consist of the state 
annihilated by the operator $\hat{a}$,
\begin{equation}
\hat{a} \vert 0 \rangle_{[0]} = 0,
\label{gr def}
\end{equation}
and the number states
\begin{equation}
\vert n \rangle_{[0]} = \frac{1}{\sqrt{n!}}
\Bigl(\hat{a}^{\dagger} \Bigr)^n \vert 0 \rangle_{[0]}.
\end{equation}
In the coordinate representation the state, $ \vert 0 \rangle_{[0]} $, 
is represented by a Gaussian wave function,
\begin{equation}
\Psi_{[0],0} (q, t) = \Bigl(\frac{1}{2 \pi u^* (t)
u (t)} \Bigr)^{1/4} \exp \Bigl[i
\frac{\dot{u}^* (t)}{2 u^* (t)} q^2 \Bigr].
\label{gr coord}
\end{equation}
Having found the eigenstates of the invariant operator, one can easily 
construct the general solution of the Schr\"{o}dinger equation with 
$\hat{H}_G$ from Eq. (\ref{sch eq2}). Note that Eq. (\ref{gr coord}) is 
the same as (\ref{Psi eq10}), which shows that our procedure indeed leads to 
the correct variational approximation of the system.
The expectation value of the Hamiltonian with respect to this state
is found to be 
\begin{equation}
{}_{[0]}\langle 0 \vert \hat{H} \vert 0 \rangle_{[0]}
= \frac{1}{2} \Bigl( \dot{u}^* \dot{u} + \omega^2 u^* u + 3\lambda \epsilon 
(u^* u)^2 \Bigr) - \frac{3}{4} \lambda \epsilon (u^* u)^2 .
\label{expec hatH}
\end{equation}

This method is also useful for solving the eigenvalue problem for the 
Hamiltonian of the time-independent anharmonic oscillator, 
in which case the parameters $\omega$ and $\tilde{\lambda}$ 
are now time-independent($\epsilon(t) = 1$). In this case Eq.(\ref{mean eq}) has a simple 
solution,
\begin{equation}
u(t) = \frac{1}{\sqrt{2\Omega_{G}}} e^{-i \Omega_{G} t},
\label{u(t) eq2}
\end{equation}
where $\Omega_{G}$ satisfies the gap equation,
\begin{equation}
{\Omega_{G}}^2 = \omega^2 + \frac{3\lambda}{2\Omega_{G}}. 
\label{eq Omega2}
\end{equation}
The state annihilated by the annihilation operator, $\hat{a}$, is 
then given by,
\begin{equation}
\Psi_{[0],0} (q, t) = \Bigl(\frac{\Omega_{G}}{\pi} \Bigr)^{1/4}
e^{-\frac{1}{2} \Omega_{G} q^2 },
\label{eq Psi2}
\end{equation}
and the expectation value of the Hamiltonian, (\ref{expec hatH}), 
reduces to
\begin{equation}
{}_{[0]}\langle 0 \vert \hat{H} \vert 0 \rangle_{[0]} 
= \frac{\Omega_{G}}{2} - \frac{3\lambda}{16{\Omega_{G}}^2 }, 
\label{expec hatH2}
\end{equation}
which coincides with the result of the Gaussian approximation of the 
anharmonic oscillator \cite{cea,lee}. This indeed shows that our method 
reproduces the well-known results of the variational approximation.

A few comments are in order. For the solution (\ref{u(t) eq2}) the terms
proportional to $\hat{a}^2$ and  $\hat{a}^{\dagger 2}$ in $\hat{H}_G$
vanish due to Eq. (\ref{eq Omega2}), which shows that the variational Gaussian 
approximation is the approximation of the interacting system by a simple
harmonic oscillator with a new frequency, $\Omega_G$.
In spite of the apparent time-dependence of the complex
function, $u(t)$, the quantum ground state,
(\ref{eq Psi2}), is time-independent
due to the choice of the solution (\ref{u(t) eq2}), which
minimizes the expectation value of the Hamiltonian, Eq. (\ref{expec hatH}).
There also exist time-dependent nontrivial solutions
to the equation (\ref{mean eq}) for the time-independent anharmonic oscillator, 
which correspond to the squeezed states in the case of the simple harmonic 
oscillator \cite{kim}. Such solutions turn out to be convenient in studying 
the time evolution of the Universe at the early stage of the inflationary 
period \cite{bak}.

Finally we remark that in the case of the simple harmonic oscillator
($\lambda = 0$), the solution (\ref{u(t) eq2}) gives the standard
vacuum, and for a time-dependent harmonic oscillator, which can be regarded as
the zero-mode of a minimal scalar field in a FRW background, it coincides
with the Bunch-Davies vacuum \cite{bunch}.

\section{Perturbative correction to the Variational Gaussian Approximation: 
Solutions of the Schr\"{o}dinger equation }

Having established the variational approximation method based on
the Liouville-Neumann equation, we now proceed to develop a
systematic method of computing the correction terms to the
variationally determined solution of the Schr\"{o}dinger equation. 
To do this we expand the annihilation
and creation type operators, $\hat{A}$ and $\hat{A}^{\dagger}$, which
satisfy the Liouville-Neumann equation (\ref{ln eq}) with the
total Hamiltonian, as a power series in the expansion parameter $\lambda$:
\begin{eqnarray}
\hat{A} = \hat{a}  + \sum_{n = 1}^{\infty}
\lambda^n \hat{B}_{(n)} = \sum_{n = 0}^{\infty} \lambda^n \hat{B}_{(n)},
\nonumber\\
\hat{A}^{\dagger} = \hat{a}^{\dagger} + \sum_{n = 1}^{\infty}
 \lambda^n \hat{B}_{(n)}^{\dagger} = \sum_{n = 0}^{\infty} \lambda^n
 \hat{B}^{\dagger}_{(n)},
\label{quant gen}
\end{eqnarray}
where $\hat{B}_{(0)} = \hat{a}$, $\hat{B}^{\dagger}_{(0)}
= \hat{a}^{\dagger}$, $\hat{a}$ and $\hat{a}^{\dagger}$
are the annihilation and creation type operators determined
in the last section, and $\hat{B}_{(n)}$ and $\hat{B}^{\dagger}_{(n)}$
are expanded as,
\begin{equation}
\hat{B}_{(n)} = \sum_{r,s}  b_{(n)}^{(r, s)} (t)
\hat{a}^{\dagger r} \hat{a}^s,
\end{equation}
with $b_{(n)}^{(r, s)}$ being the $c$-number function of time [Note that 
the operators $\hat{B}_{(n)}$ and $\hat{B}^{\dagger}_{(n)}$ are dependent on 
$\epsilon(t)$ through $b_{(n)}^{(r, s)}$].
Substituting Eq. (\ref{quant gen}) into the Liouville-Neumann
equation (\ref{ln eq}), and by comparing the coefficients of
$\lambda^n$, one obtains the equation,
\begin{equation}
\frac{\partial}{\partial t} \hat{B}_{(n)} = i
[ \hat{B}_{(n)}, \hat{H}_G ]
+ i [ \hat{B}_{(n-1)}, \hat{H}' ],
\label{nth eq}
\end{equation}
where $\hat{B}_{(-1)} = 0$.

We note that any product of the operators
that satisfy Liouville-Neumann equation
also satisfies the Liouville-Neumann equation. Thus,
the operator $\hat{a}^{\dagger r} \hat{a}^s$ satisfies the equation,
\begin{equation}
\frac{\partial}{\partial t} \Bigl(\hat{a}^{\dagger r} \hat{a}^s \Bigr)
= i [ \hat{a}^{\dagger r} \hat{a}^s, \hat{H}_G],
\end{equation}
and Eq. (\ref{nth eq}) reduces to
\begin{equation}
\sum_{r,s} \Bigl(\frac{\partial}{\partial t} 
b_{(n)}^{(r, s)} \Bigr) \hat{a}^{\dagger r} \hat{a}^s
= i \sum_{r,s} b_{(n-1)}^{(r, s)} [
\hat{a}^{\dagger r} \hat{a}^s, \hat{H}'].
\label{operator eq2}
\end{equation}
Note that the time derivative operator in the left hand side
of Eq.(\ref{operator eq2}) acts only on the coefficients
$b_{(n)}^{(r, s)}$. This shows that we can find $\hat{B}_{(n)}$
by treating $\hat{a}$ and $\hat{a}^{\dagger}$ as if they are
time-independent:
one can integrate the equation (\ref{operator eq2}) to get
\begin{eqnarray}
\hat{B}_{(1)} &=& i \int [ \hat{a}, \hat{H}'],
\nonumber\\
&\vdots&
\nonumber\\
\hat{B}_{(n)} &=& i^n \int \cdots \int
[[[[\hat{a}, \hat{H}'], \hat{H}'], \cdots],
\hat{H}'],
\end{eqnarray}
where the integrals are ordered from the inner
bracket to the outer bracket, i.e., for example,
\begin{equation}
\hat{B}_{(2)} = i^2 \int_{0}^{t} dt' \int_{0}^{t'} dt''
[[\hat{a}, \hat{H}' (t'')], \hat{H}' (t')].
\end{equation}

The state annihilated by $\hat{A}$, 
\begin{equation}
\hat{A} \vert 0 \rangle = 0,
\label{vacuum}
\end{equation}
can also be expanded as
\begin{equation}
\vert 0 \rangle =
 \sum_{k = 0}^{\infty} \lambda^k \hat{Q}_{(k)} \vert 0
 \rangle_{[0]},
 \label{vacuumstate2}
\end{equation}
where $\vert 0 \rangle_{[0]}$ is the variationally determined
state, Eqs. (\ref{gr def}) and (\ref{gr coord}), 
$\hat{Q}_{(0)} = 1$ and
$\hat{Q}_{(k)}$ is an operator to be determined.
Substituting (\ref{vacuumstate2}) into (\ref{vacuum}) we obtain
\begin{equation}
\hat{A} \vert 0 \rangle =
\sum_{k, l = 0}^{\infty} \lambda^{k+l}
\hat{B}_{(l)} \hat{Q}_{(k)} \vert 0 \rangle_{[0]} = 0.
\label{hatA eq2}
\end{equation}
For Eq. (\ref{hatA eq2}) to be satisfied, each coefficient
of $\lambda^n$ must vanish:
\begin{equation}
\sum_{k, l = 0}^{k+l = n}
\hat{B}_{(l)} \hat{Q}_{(k)} \vert 0 \rangle_{[0]} = 0,
\end{equation}
which can be rewritten as
\begin{equation}
\hat{B}_{(0)} \hat{Q}_{(n)} \vert 0 \rangle_{[0]}
= \hat{a}\hat{Q}_{(n)} \vert 0 \rangle_{[0]}
= - \sum_{k = 0}^{n-1}
\hat{B}_{(n-k)} \hat{Q}_{(k)} \vert 0 \rangle_{[0]}.
\label{hatBQ eq2}
\end{equation}
By operating $\hat{a}^{\dagger}$ on both sides of Eq. (\ref{hatBQ eq2}),
one finally obtains the relation,
\begin{equation}
\hat{Q}_{(n)} \vert 0 \rangle_{[0]} =
- \frac{1}{\hat{N}}\hat{a}^{\dagger}
 \sum_{k = 0}^{n-1}
\hat{B}_{(n-k)} \hat{Q}_{(k)} \vert 0 \rangle_{[0]},
\label{hatQ eq2}
\end{equation}
where $\hat{N}$ is the number operator, $\hat{a}^{\dagger} \hat{a}$. 
Note that the sum on $k$ in the right hand side of Eq. (\ref{hatQ eq2}) 
runs only up to $(n-1)$,
and thus we can construct the operator $\hat{Q}_{(n)}$ by
applying Eq. (\ref{hatQ eq2}) iteratively, starting from the first
non-trivial contribution. Substituting (\ref{hatQ eq2}) into 
(\ref{vacuumstate2}) we obtain the state annihilated by $\hat{A}$ 
as a perturbation series, from which solutions of the Schr\"{o}dinger equation 
can be constructed from Eq. (\ref{sch eq2}).

\section{Eigenvalue problem of the time-independent Anharmonic Oscillator}

To illustrate the effectiveness of our approximation method we now compute the 
ground state of the time-independent anharmonic oscillator described by the 
Hamiltonian (\ref{anhar}) with time-independent parameters $\omega$ and $\tilde{\lambda}$ 
$(\epsilon(t) = 1)$.

For the time-independent systems we are interested in the ground state of the 
system, which is an eigenstate of the Hamiltonian of the system. In the 
introduction we have shown that the solution of the Schr\"{o}dinger equation
(\ref{sch eq}) can be written as a linear combination of the eigenstates, 
$\vert e_n, t \rangle$, of the invariant operator $\hat{I}$, which do not necessarily
coincide with the eigenstates of the Hamiltonian.
If the eigenstates are time-independent, 
$\frac{\partial}{\partial t} \vert e_n, t \rangle = 0$, however, one can easily 
show from Eqs. (\ref{ln eq}) and (\ref{eigen}) that such states are the simultaneous 
eigenstates of $\hat{I}$ and $\hat{H}$:
\begin{equation}
[\hat{H} , \hat{I}] \vert e_n, t \rangle = 0,
\label{commutation eq} 
\end{equation}
and 
\begin{equation}
\hat{H} \vert e_n, t \rangle = h_n \vert e_n, t \rangle,
\label{eigenvalueH eq3} 
\end{equation}
from the non-degeneracy of $\hat{I}$. It follows from this that the state 
annihilated by $\hat{A}$, Eq. (\ref{vacuumstate2}), is indeed the ground state of 
the time-independent system if it is time-independent.

We have shown in section II that the state annihilated by $\hat{a}$ for the 
time-independent anharmonic oscillator, Eq. (\ref{eq Psi2}), is time-independent.
This fact guarantees that the state (\ref{eq Psi2}) is indeed the variationally 
approximated ground state and the expectation value of $\hat{H}$ with respect to
this state is the approximate ground state energy of the system. And the state
annihilated by $\hat{A}$, Eq. (\ref{vacuumstate2}), for the time-independent 
system is the perturbatively corrected ground state if it is time-independent.

The ground state energy of the time-independent anharmonic oscillator can be 
obtained from the ground state (\ref{vacuumstate2}),
\begin{equation}
E_g = \sum_{n = 0}^{\infty} {\lambda}^{n} E_{g,(n)},
\label{ground state energy}
\end{equation}
where 
\begin{eqnarray}
E_{g,(0)} &=& \frac{\Omega_G}{2} - \frac{3\lambda}{16{\Omega_{G}}^2 }
\nonumber\\
E_{g,(1)} &=& {}_{[0]}\langle 0 \vert \hat{H}' \vert 0 \rangle_{[0]} = 0 \\
\label{ground energy2}
E_{g,(n)} &=& {}_{[0]}\langle 0 \vert \hat{H}' {\hat{Q}}_{(n-1)} \vert 0 \rangle_{[0]}
- \sum_{k=0}^{n-1} E_{g,(k)} {} 
{}_{[0]}\langle 0 \vert {\hat{Q}}_{(n-k)} \vert 0 \rangle_{[0]}, \quad n\ge 2. \nonumber
\end{eqnarray}
Note that, due to the property of the variational perturbation theory, the 
first order correction to the energy eigenvalue vanishes.

To show the time-independence of the quantum ground state,
and to compare with the standard method,
it is convenient to write the operators in Eq. (\ref{cr-an}) as
\begin{equation}
\hat{a} = e^{i \Omega_G t} \hat{a}_0, ~ \hat{a}^{\dagger} = e^{-i
\Omega_G t} \hat{a}^{\dagger}_0,
\label{new cr-an}
\end{equation}
where $\hat{a}_0$ and $\hat{a}^{\dagger}_0$ are the time-independent, standard
annihilation and creation operators. It is manifest that
the operators in Eq. (\ref{new cr-an}) satisfy the
Liouville-Neumann equation (\ref{ln eq}) with the truncated
Hamiltonian which can be written as
\begin{equation}
\hat{H}_G = \Omega_G \Bigl(\hat{a}^{\dagger} \hat{a} +
\frac{1}{2} \Bigr) = \Omega_G \Bigl(\hat{a}^{\dagger}_0 \hat{a}_0 +
\frac{1}{2} \Bigr).
\end{equation}
The remaining perturbation Hamiltonian becomes
\begin{equation}
\hat{H}' = \frac{1}{4 \Omega_G^2} \Bigl(\frac{1}{4} \hat{a}^{\dagger 4}_0
+ \hat{a}^{\dagger 3}_0 \hat{a}_0 + \frac{3}{2} \hat{a}^{\dagger
2}_0 \hat{a}^2_0  + \hat{a}^{\dagger}_0 \hat{a}^3_0
+ \frac{1}{4} \hat{a}^4_0 \Bigr).
\end{equation}
From now on we shall work with both the time-dependent
operators $\hat{a}$, $\hat{a}^{\dagger}$ and time-independent
ones $\hat{a}_0$, $\hat{a}^{\dagger}_0$, whenever it is more convenient.

We now want to compute the operators $\hat{A}$ and $\hat{A}^{\dagger}$
and the vacuum state. It is more convenient to work with the
time-independent, standard operators, in terms of which
Eq. (\ref{operator eq2}) reads 
\begin{equation}
\sum_{r,s} e^{-i (r-s) \Omega_G t} \frac{\partial}{\partial t} \Bigl(
b_{(n)}^{(r, s)} \Bigr) \hat{a}^{\dagger r}_0 \hat{a}^s_0
= i \sum_{r,s} e^{-i (r-s) \Omega_G t} b_{(n-1)}^{(r, s)} [
\hat{a}^{\dagger r}_0 \hat{a}^s_0, \hat{H}'].
\end{equation}
It is straightforward to compute the first-order correction term of the 
annihilation operator $\hat{A}$,
\begin{equation}
\hat{A}_{[1]} = e^{i \Omega_G t}
\Biggl[ \hat{a}_0 + \frac{\lambda}{(2 \Omega_G)^2}
\Biggl\{\frac{1}{4 \Omega_G} \hat{a}^{\dagger 3}_0
+ \frac{3}{2 \Omega_G} \hat{a}^{\dagger 2}_0
\hat{a}_0 + (3 i t + c_0) \hat{a}^{\dagger}_0 \hat{a}^2_0
- \frac{1}{2 \Omega_G} \hat{a}^{3}_0 \Biggr\}\Biggr],
\end{equation}
where $c_0$ is an integration constant.
$\hat{A}^{\dagger}_{[1]}$ is the Hermitian conjugate of
$\hat{A}_{[1]}$.
We thus find, from Eq. (\ref{vacuumstate2}), the first order vacuum state,
\begin{eqnarray}
\vert 0 \rangle_{[1]} &=& \vert 0 \rangle_{[0]} +
\lambda \hat{Q}_{(1)} \vert 0 \rangle_{[0]}
\nonumber\\
&=& \Bigl[ 1 - \frac{\lambda}{4^3 \Omega_G^3}
\hat{a}_0^{\dagger 4} \Bigr] \vert 0 \rangle_{[0]}.
\label{38}
\end{eqnarray}
and the corresponding vacuum energy,
\begin{equation}
E_g = \frac{\Omega_G}{2}- \frac{3\lambda}{16{\Omega_{G}}^2 }
- \frac{3 \lambda^2}{128 \Omega_G^5} + {\cal O} (\lambda^3),
\end{equation}
which is the same as that obtained by other variational perturbation theories \cite{cea,lee}. 
Note that the first order
vacuum state $\vert 0 \rangle_{[1]}$ is time-independent as
it should be. 

We now consider an interesting algebraic structure that the first order variational perturbation
of the anharmonic oscillator exhibits.
With the specific choice of the integration constant, $c_0 = \frac{3}{4}$,
we find that to the order of $\lambda$ the full Hamiltonian
can be expressed as,
\begin{equation}
\hat{H} = \Omega_G \hat{A}_{[1]}^{\dagger} \hat{A}_{[1]}
+ \frac{\Omega_G}{2}- \frac{3 \lambda}{16{\Omega_{G}}^2 } + {\cal O} (\lambda^2),
\label{eq 48}
\end{equation}
and the commutation relation between $\hat{A}$ and $\hat{A}^+$ becomes,
\begin{eqnarray}
[\hat{A}_{[1]}, \hat{A}_{[1]}^{\dagger} ] &=&
1 + \frac{3 \lambda}{4 {\Omega}^2_{G}} \hat{a}^{\dagger}_0 \hat{a}_0
+ {\cal O} (\lambda^2)
\nonumber\\
&=& 1 + \frac{3 \lambda}{4 {\Omega}^2_{G}} \hat{A}_{[1]}^{\dagger}
\hat{A}_{[1]} + {\cal O} (\lambda^2).
\label{def alg}
\end{eqnarray}
The algebraic structure of Eq. (\ref{eq 48})
for the Hamiltonian explains why the state $\vert 0 \rangle_{[1]}$
annihilated by $\hat{A}_{[1]}$ is the vacuum state for the full Hamiltonian
up to the order of $\lambda$. It should be noted that when
the invariant operators and the full Hamiltonian are truncated
to the order of $\lambda^2$, the truncated Hamiltonian (\ref{eq 48}) has 
a factorized form and reminds us of a possible underlying
supersymmetry and a deformed commutator (\ref{def alg}).
The deformation parameter is greater than one, so
the deformed algebra does not interpolate between bosons and fermions.
However, the deformed algebra can be used conveniently to find the excited
quantum states.

\section{Discussions and Conclusion}
We have developed a variational perturbation theory based on the Liouville-Neumann 
equation, which enables one to compute the order-by-order correction terms to the 
variationally approximated wave functions of the time-dependent systems. The method 
is based on the observation that the variationally determined approximate eigenstates 
of the Hamiltonian of an interacting system are the exact eigenstates of a new quadratic
(or, a soluble) Hamiltonian, and on the fact that the solution of the Schr\"{o}dinger equation 
for a time-dependent harmonic oscillator can be represented by eigenstates of an invariant 
operator satisfying the Liouville-Neumann equation (\ref{ln eq}). We thus write the Hamiltonian 
of an interacting system as a function of annihilation and creation type operators $\hat {a}$ and 
$\hat {a}^+$, which are defined as linear functions of dynamical variables with an arbitrary complex
function $u(t)$ introduced. We then truncate the Hamiltonian up to quadratic terms in $\hat {a}$ and 
$\hat {a}^+$, and require $\hat {a}$ and $\hat {a}^+$ to satisfy the Liouville-Neumann equation with this 
truncated Hamiltonian $\hat H_G$. This requires the function $u(t)$ to satisfy a gap equation, which 
is exactly the same as that appearing in the Gaussian approximation method as shown in section II. 
This constitutes the Liouville-Neumann approach to the variational Gaussian approximation method.

To develop a perturbation theory based on this variational approximation, we expand the annihilation
and creation type operators $\hat {A}$ and $\hat {A}^+$, for the full interacting Hamiltonian, as power 
series in the coupling constant $\lambda$. By requiring the operators $\hat {A}$ and $\hat {A}^+$ to satisfy 
the Liouville-Neumann equation with the total Hamiltonian, these operators $\hat {A}$ and $\hat {A}^+$ can 
be determined to each order in the coupling constant $\lambda$. By constructing the state annihilated by 
the operator $\hat {A}$, we obtain the quantum state satisfying the Schr\"{o}dinger equations to each 
order in $\lambda$. Since this method enables one to construct wave functions of time-dependent 
systems, it can be conveniently used to study the time-evolution of time-dependent systems such as 
time-evolution of scalar field in the expanding Universe.

To find the eigenstates of time-independent Hamiltonian, one needs one more requirement to the 
procedure explained above. For the eigenstates of the invariant operator $\hat I$ to be the 
simultaneous eigenstates of the Hamiltonian, the eigenstates must be time-independent as shown 
in the last section. We have used this method to compute the ground state of the time-independent 
anharmonic oscillator to the first order in $\lambda$, and have shown that the result agrees 
with that of other variational perturbation theories. To this order of the perturbation the 
Hamiltonian is found to be factorized as in the simple harmonic oscillator with the annihilation 
and creation operator $\hat {A}$ and $\hat {A}^+$, which satisfy the deformed algebra (\ref{def alg}). 
This interesting algebraic structure can be used to construct all the excited states of the 
system. This possibility will be explored elsewhere \cite{bak2}.

Although we have considered the quantum mechanical anharmonic oscillator for simplicity, 
this method can easily be generalized to the case of quantum field theory.

\section*{Acknowledgments}
This work was supported by the Basic Science Research Institute Program,
Korea Ministry of Education under Project No. 98-015-D00054, 98-015-D00061, 98-015-D00129,
and by the Center for Theoretical Physics, Seoul National University.
SPK was also supported by the Non-Directed Research Fund, the
Korea Research Foundation, 1997, JHY  by the Korea Science and
Engineering Foundation under Project No. 97-07-02-02-01-3, and DB by KOSEF Interdisciplinary 
Research Grant 98-07-02-07-01-5.

\end{document}